\documentclass[useAMS,usenatbib]{mn2e}

\usepackage{graphicx}
\usepackage{natbib}

\title[The binary fraction in five open clusters]{The fraction of binary systems in the core of five Galactic open clusters\thanks{Based on data obtained from the ESO Science Archive Facility and Isaac Newton Group Archive}}
\author[Sollima et al.]{A. Sollima$^{1}$\thanks{E-mail:
asollima@iac.es (AS)}, J. A. Carballo-Bello$^{1}$, G. Beccari$^{2}$, F. R.
Ferraro$^{3}$, F. Fusi Pecci$^{4}$ and\\
\newauthor
B. Lanzoni$^{3}$\\
$^{1}$Instituto de Astrofisica de Canarias, c/via Lactea s/n, La Laguna,
38205-E, Spain\\
$^{2}$European Space Agency, Space Science Department, Keplerlaan 1, Noordwijk,
2200 AG, The Netherlands\\
$^{3}$Dipartimento di Astronomia, Universit\`a di Bologna, via Ranzani 1,
Bologna, 40127-I, Italy\\
$^{4}$INAF Osservatorio Astronomico di Bologna, via Ranzani 1,
Bologna, 40127-I, Italy}

\begin{document}

\date{Accepted 2009 August ??; Received 2009 August ??; in original form
2009 July ??}

\pagerange{\pageref{firstpage}--\pageref{lastpage}} \pubyear{2009}

\maketitle

\label{firstpage}

\begin{abstract}
We used deep wide field photometric observations to derive the 
fraction of binary systems in a sample of five high-latitude Galactic open 
clusters. 
By analysing the color distribution of Main Sequence stars we derived the
minimum fraction of binary systems required to reproduce the observed 
color-magnitude diagram morphologies. 
We found that all the analysed clusters contain a minimum binary fraction 
larger than 11\% within the core radius. 
The estimated global fractions of binary systems range from 35\% to 70\% 
depending on the cluster.
The comparison with homogeneous estimates performed in globular clusters 
indicates that open clusters hold a significantly higher fraction of binary 
systems, as predicted by theoretical models and N-body simulations.
A dependence of the relative fraction of binary systems on the cluster 
mass has been detected, suggesting that the binary disruption within the cluster 
core is the dominant process that drives the fraction of binaries in stellar
systems.
\end{abstract}


\section{Introduction}
\label{intro}

The study of the binary star population in stellar systems represents an
important field of research of stellar astrophysics. 
Binary stars are a unique tool to determine crucial information about a 
variety of stellar properties.
Moreover, they play a key role in the dynamical evolution of stellar systems and 
stellar populations studies. 
In collisional systems (like stellar clusters) binaries provide the 
gravitational fuel that can delay and eventually stop and reverse the process of 
gravitational collapse (see Hut et al. 1992 and references therein).
Furthermore, the evolution of binaries in star clusters can produce peculiar 
objects of astrophysic interest like blue stragglers, cataclysmic variables, 
low-mass X-ray binaries, millisecond pulsars, etc. (see Bailyn 1995 and 
reference therein). 
Finally, the binary fraction is a key ingredient in dynamical models to
study the evolution of galaxies and stellar systems in general (Zhang et al. 2005).

At odds with the Galactic field, several processes determine 
the relative frequency of binary systems in stellar clusters. 
In fact, binaries are continuously formed and destroyed during the evolution of
the cluster as a result of the ever continuing interactions between binaries 
and single stars. The scenario is further complicated by the dynamical 
evolution of the cluster (mass segregation, evaporation,
etc.) that acts on binary and single stars in a different way and produces radial
gradients in the frequency of binaries.
For these reasons, the theoretical modelling of the dynamics of stellar systems
including the effect of binaries is still an open challenge (Portegies Zwart,
McMillan \& Makino 2007; Ivanova et al. 2005; Hurley, Aarseth \& Shara 2007; Sollima 2008).

%From the observational point of view, the binary 
%fraction in the Galactic field and in stellar clusters has been estimated by 
%means of extensive radial velocity variability surveys (Abt et al. 1970; 
%Duquennoy \& Mayor 1991; Latham 1996; Halbwachs et al. 2003).
%However, the nature of this method leads to intrinsic observational biases and 
%a low detection efficiency.
%In single stellar populations (like many open and globular clusters), 
%a powerful method to estimate the fraction of binaries is
%to study the number of stars displaced in the secondary Main-Sequence (MS) in 
%color-magnitude diagrams (CMD, Rubenstein \& Bailyn 1997).
%This method represents a more efficient statistical approach and  
%does not suffer of strong selection biases.  
%
From the observational point of view, until recent years, the binary fraction 
has been estimated only in few individual 
globular clusters (GCs) (Romani \& Weinberg 1991; Bolte 1992; 
Rubenstein \& Bailyn 1997; Bellazzini et al. 2002; Clark, 
Sandquist \& Bolte 2004; Zhao \& Bailyn 2005).
These studies argued for a deficiency of binary stars in GCs 
compared to the field (Pryor et al. 1989; Hut 1992; Cote et al. 1996).
More recently, we investigated the fraction of binaries 
in a sample of thirteen low-density GCs (Sollima et al. 2007).
This study revealed that these GCs hold a fraction of
binaries ranging from 10\% to 50\% depending on the cluster.
In such analysis, an anti-correlation with the cluster age has been also noticed.
Milone et al. (2008) enlarged the sample of analysed clusters and showed an even
stronger anti-correlation between binary fraction and cluster luminosity (mass).
This last correlation is predicted by theoretical models as a
consequence of the similar dependence of the cluster mass and of the efficency of the binary
destruction process on the cluster density and velocity dispersion (Sollima
2008).
Unfortunately, the sparse number of analysed objects together with the small 
range of parameters covered by the sample does not allow a firm conclusion on 
this issue.

Open clusters (OCs) represents an important group of objects to study the
frequency of binary systems. 
They are indeed both less massive and younger than GCs, covering
a range of structural parameters where homogeneous determinations of binary
fractions are still missing. 
Moreover, their proximity and low density makes the determination of the binary
frequency in these stellar systems particularly easy.
Estimates of the binary fraction in individual open OCs have been provided
by several authors (Sandhu, Pandey \& Sagar 2003; Bica \& Bonatto 2005 and references therein).
However, it is difficult to interpret the results obtained by these authors in a
global picture because of the different assumption made in these works.
A recent homogeneous analysis of the fraction of binaries in a sample of six 
OCs have been presented by Sharma et al. (2008). 
Nevertheless, a direct comparison of these last results with those obtained for GCs 
is not possible since these authors {\it i)} assume a
significantly different distribution of mass-ratios and {\it ii)} measure the 
fraction of binaries over the entire cluster extent.

In this paper we present an estimate of the binary fraction in the core of five
high-latitude OCs. We used a set of archive images obtained with
the Wide Field Imager (at the ESO2.2m telescope), Wide Field Camera (Isaac 
Newton Telescope) and 1.5m Danish telescope cameras.   

In \S 2 we describe the observations and the data reduction techniques. 
In \S 3 the adopted method to determine the fraction of binary systems is 
presented. In \S 4 we derived the binary
fractions in our target clusters. \S 5 is devoted to the analysis of the
correlations between the binary fractions measured in our sample of OCs and GCs
and the main cluster's physical parameters.
Finally, we summarize and discuss our results in \S 6.

\section{Observations and Data reduction}
\label{reduct}

The photometric data-set consists of a set of wide-field images of a sample of
five OCs. 
The target clusters have been selected on the basis of the following criteria:
\begin{itemize}
\item A high Galactic latitude ($b>15^{\circ}$) in order to limit the field
contamination;
\item An absolute visual magnitude $M_{V}<-2.5$ in order to detect a significant
sample of cluster stars.
\end{itemize}
Five cluster passed these criteria namely NGC188, NGC2204, NGC2243,
NGC2420 and NGC2516.
In Table 1 the main physical parameters of the above target clusters are listed.
The age ($t_{9}$), the metallicity ([Fe/H]) and the Galactic latitude ($b$)
are from the WEBDA database (Mermilliod \& Paunzen 2003) while the V absolute magnitude ($M_{V}$) 
from Lata et al. (2002)\footnote{For NGC2516, which is not included in 
the list of Lata et al. (2002), we adopted the value of $M_{V}$ obtained by
Battinelli \& Capuzzo-Dolcetta (1991).}. 

For each cluster we retrived all the available exposures in the B and V bands 
from the ESO and ING Science Archives with the WFI and WFC cameras, respectively. 
WFI@ESO2.2m frames of NGC2204 and NGC2516 and 
WFC@INT images of NGC188 and NGC2420 have been retrived. 
Images cover an area of $34\arcmin\times33\arcmin$ around the 
center of these clusters. Each cluster has been centered in one of the chips 
of the camera, allowing to sample the entire cluster extent.
In addition, small field ($3\arcmin\times5\arcmin$) images obtained at the
1.5m Danish telescope have been used to study the core of NGC2243. 

After applying the standard bias and flat-field correction, the photometric
analysis has been performed on the pre-reduced images
using the SExtractor photometric package (Bertin \& Arnouts 1996). 
Given the small star density in these clusters ($\leq 0.03~stars~arcsec^{-2}$ at
$V<20$), 
crowding does not affect the aperture photometry, allowing to properly estimate 
the magnitude of stars. 
For each star we measured the flux contained within a radius r$\sim$FWHM from the star center. 
After applying the correction for exposure time, airmass and nominal infinite aperture, 
instrumental magnitudes have been transformed into
the absolute Johnson system by using a set of standard stars from the Landolt
(1992) list observed during each observing run. 

Previous photometric analysis are present in the literature for all the clusters
analysed in this work.
Our photometry has been compared with the photometric catalog already published
(Krusberg \& Chaboyer 2006 for NGC188; Kassis et al. 1997 for NGC2204; 
Bonifazi et al. 1990 for NGC2243; Sharma et al. 2006 for NGC2420 and Jeffries,
Thurston \& Hambly 2001 for NGC2516). 
The mean magnitude differences found are always smaller than 0.02 mag in both passbands,
consistent with no systematic offset.
Optimal astrometric solutions have been obtained through cross-correlation with a catalog suitable 2MASS astrometric standard stars. The internal accuracy of the astrometry has been found to be $<0.2\arcsec$. 

\begin{figure*}
 \includegraphics[width=12.cm]{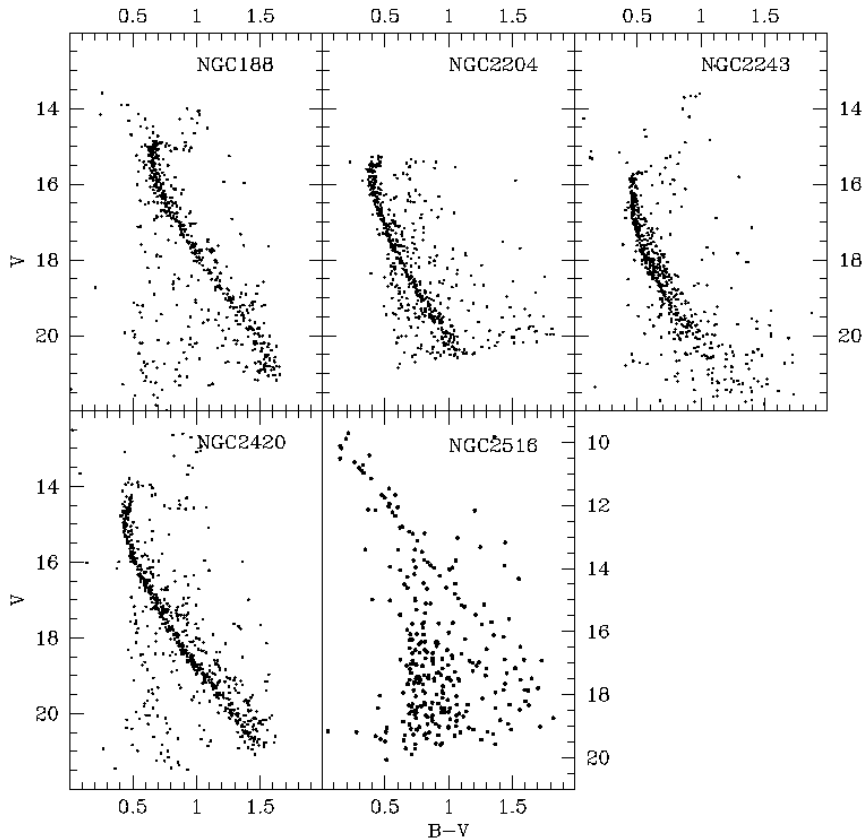}
\caption{$V, B-V$ CMDs of the target open clusters. Only stars within 
5$\arcmin$ from the cluster centers are shown.}
\label{cmd}
\end{figure*}

Fig. \ref{cmd} shows the ($V, B-V$) CMDs of the 
5 OCs in our sample. Only stars within 5$\arcmin$ from the cluster center are
shown. 
The CMDs sample the cluster
population from the sub-giant branch down to 4-5 magnitudes below the MS turn-off.
In all the target clusters the binary sequence is well defined and distinguishable 
from the cluster's MS. 

\begin{table}
\label{riass}
 \centering
  \caption{Main physical parameters of the target clusters}
  \begin{tabular}{@{}lccccr@{}}
  \hline
   Name     & $t_{9}$ & [Fe/H] & $b$     & $M_{V}$ & $r_{c}$\\
            & (Gyr)   &        & (deg)   &         & ($\arcmin$)\\
 \hline
 NGC 188    & 4.30    & -0.02  &  22.384 & -2.86 & 3.79\\
 NGC 2204   & 1.60    & -0.33  & -16.107 & -4.65 & 4.15\\
 NGC 2243   & 3.80    & -0.44  & -18.014 & -2.67 & 2.00\\
 NGC 2420   & 1.20    & -0.26  &  19.634 & -3.44 & 2.31\\
 NGC 2516   & 0.33    & +0.06  & -15.865 & -4.82 & 5.00\\
 \hline
\end{tabular} 
\end{table}

\section{Method}
\label{method}

To determine the fraction of binaries in our sample of OCs we analysed
the distribution of stars in the CMD displaced above the cluster Main-Sequence (MS) 
following the method described in Sollima et al. (2007).
Any binary system is indeed seen as a single star with a flux 
equal to the sum of the fluxes of the two components. 
This effect produces a systematic overluminosity of these objects and a shift 
in color depending on the magnitudes of the two components in each passband. 
In a simple stellar population the luminosity of a MS star is univocally 
connected with its mass according to a mass-luminosity relation.  
The magnitude of a given binary system can be written as:
$$m_{sys}= -2.5~log(F_{M_{1}}+F_{M_{2}})+c$$
$$=m_{M_{1}}-2.5~log(1+\frac{F_{M_{2}}}{F_{M_{1}}})$$ 
Where $M_{1}$ and $M_{2}$ are the mass of the most massive (primary) and less
massive (secondary) component, respectively, and $F_{M}$ the flux emitted in a
given passband, 
The quantity $\frac{F_{M_{2}}}{F_{M_{1}}}$ depends on the mass ratio of the two
component ($q=\frac{M_{2}}{M_{1}}$). According to the definition of $M_{1}$ and
$M_{2}$ given above, the parameter $q$ is comprised in the range $0<q<1$. 
When q=1 (equal mass binary) the binary system will
appear $-2.5~log(2)\sim0.752$ mag brighter than the primary component.
Conversely, when $q$ approaches small values the ratio
$\frac{F_{M_{2}}}{F_{M_{1}}}$ approaches zero, producing a negligible
shift in the CMD with respect to the primary star.
Following these considerations, only binary systems with values of $q$ larger than a minimum value ($q_{min}$) 
are unmistakably distinguishable from single MS stars. 
For this reason, only a lower limit to the binary fraction can be directly derived
without assuming a specific distribution of mass-ratios $f(q)$.
The value of $q_{min}$ depends on the photometric accuracy of the data.

Further complications are due to two important effects: {\it i)}
blended source contamination and {\it ii)} field star contamination.
Moreover, since our analysis is addressed to the measure of the binary fraction
in the core of the target OCs, an accurate determination of the core radius is
also necessary.

In order to study the relative frequency of binary systems in our target
clusters we followed two different approaches:
\begin{itemize}
\item We derived the minimum number of binary systems by considering only the
fraction of binary systems with large mass-ratio values ($q>q_{min}$);
\item We estimated the {\it complete} binary fraction by assuming a
given $f(q)$ and comparing the simulated CMDs with the observed ones.
\end{itemize}

In the following we will refer to the binary fraction $\xi$ as the ratio between 
the number of binary systems whose primary star has a mass comprised in a 
given mass range ($N_{b}$) and the number of cluster members in the same mass
range ($N_{tot}=N_{MS}+N_{b}$).

The procedure to derive an accurate estimate of the binary fraction in each
target cluster can be schematically summarized as follows:
\begin{itemize}
\item{We derived the projected density profile of the cluster and estimated its core radius (\S \ref{prof_sec});}
\item{We defined on the CMD of each cluster two regions which divide the cluster
population in two samples: the $MS~sample$ and the $binary~sample$ (\S
\ref{box_sec});}
\item{We simulated a synthetic population of single and binary stars that have
been added on the original frames and used to quantify the impact of blends,
completeness and photometric errors (\S \ref{blend});}
\item{We calculated the frequency of field stars contaminating the $MS~sample$ 
and the $binary~sample$ (\S \ref{field});}
\item{We search for the fraction of binaries that
reproduces the observed ratio of $MS~sample$ and the $binary~sample$ stars 
inside the cluster core radius (\S \ref{frac}).}
\end{itemize}

In the following sections the adopted procedures to perform the above tasks are
presented.

\subsection{Density profile and core determination}
\label{prof_sec}

OCs are collisional systems whose mean relaxation time is generally smaller
than their age.
As a consequence, the process of mass segregation is efficient in these systems
producing radial gradients in 
the distribution of stars with different masses.
In particular, massive stars tend to sink into the cluster core after a
time-scale comparable with the cluster relaxation time.
Being bound systems, binary stars dynamically behave like a single star with a 
mass equal to the sum of the masses of the two components. Hence, being on average more
massive than sigle stars, they will populate preferentially the most
internal regions of the cluster.

To compare the binary fraction in clusters with different structural properties,
is therefore necessary to refer the analysis to a characteristic radius.
Here we estimated the fraction of binaries in the core of the clusters of our
sample.

\begin{figure*}
 \includegraphics[width=12.cm]{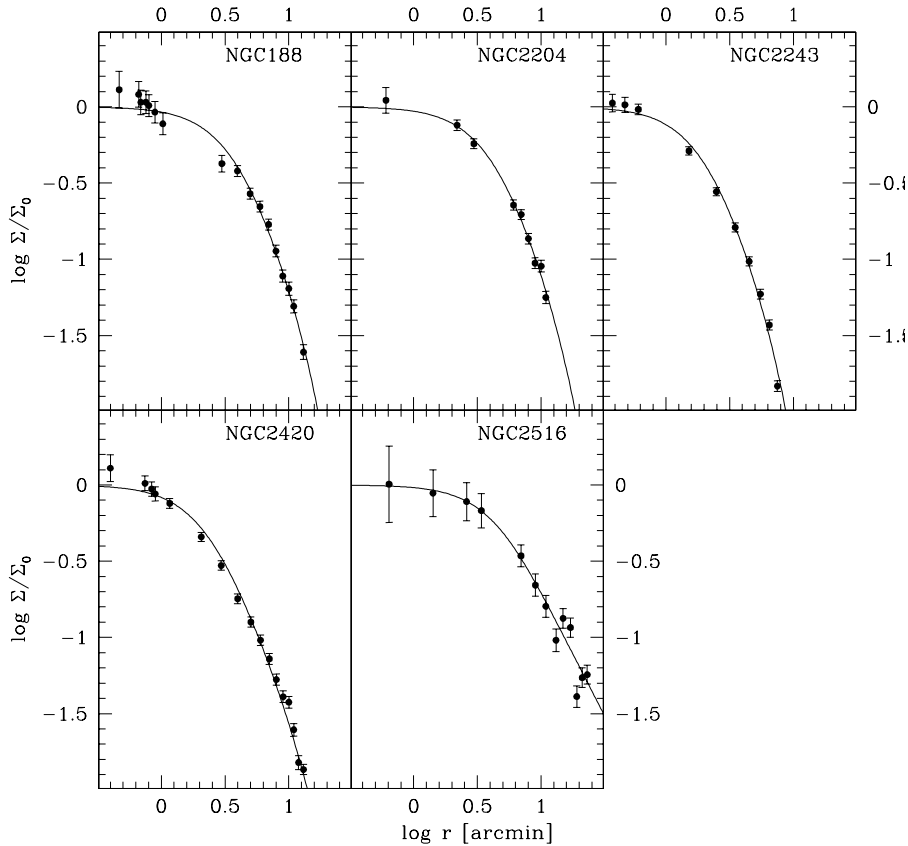}
\caption{Projected surface density of the target open clusters. The best-fit King
(1966) models are overplotted.}
\label{den}
\end{figure*}

To determine the dimension of the cluster's core, we selected for each cluster 
all the stars with a color difference from the MS mean ridge line
smaller then 3 times the color photometric error corresponding to their magnitude
and a magnitude which ensures a completeness factor $\phi>50$\% (estimated from
artificial stars experiments; see Sect.\ref{blend}).
The field of view has been therefore divided in concentric annuli of variable
width and the projected surface density of MS stars has been calculated.
For NGC2243, whose observations covers only a small portion of the cluster
extent, we used the photometric catalog by Kaluzny, Krezminski \& Mazur (1996) to
calculate the cluster density profile.  
The core radius for each cluster has been obtained by best-fitting its density
profile with a suitable King (1966) model.  
As can be noted in Fig. \ref{den}, in some cluster there are outliers whose
density estimate stray from the best-fit King model. 
This is a consequence of the low density of these clusters which
makes critical the density estimation. To quantify the uncertainties in the
derived core radii we carried out an iterative procedure: in each iteration a
half of the sample points have been randomly extracted and a King model has been 
fitted to determine the core radius. 
The standard deviation of the whole set of determinations gives an estimate of 
the overall uncertainty in the estimated core radius.
This procedure applied to our sample of clusters indicates an accuracy always
better than 10\% 
. Such an uncertainty has only a negligible effect on the binary
fraction determination.

The projected densities measured for the target OCs are shown in Fig.
\ref{den}. The best-fit King(1966) models are overplotted and the corresponding
core radii are listed in Table 1.
The derived core radii are in good agreement with the estimates available in the
literature (Bonatto \& Bica 2005; Bonatto, Bica \& Santos 2005; 
Sharma et al. 2006). 

\subsection{Sample definition}
\label{box_sec}

\begin{figure}
 \includegraphics[width=8.7cm]{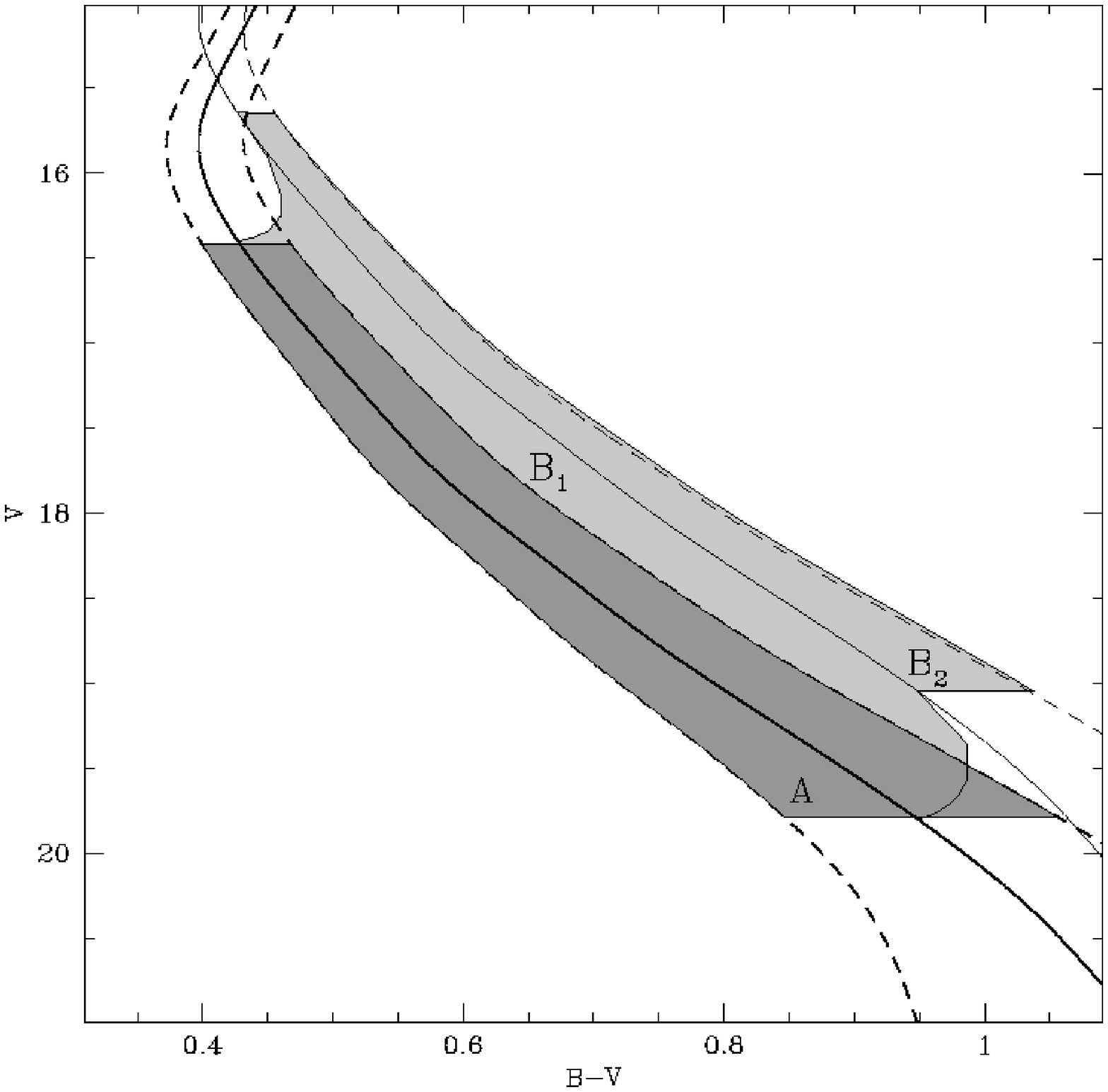}
\caption{Selection boxes used to select the $MS sample$ (dark grey area) and the 
$binary~sample$ (grey area). The solid
thick line marks the MS mean ridge line, the solid thin line marks the equal-mass
binary sequence, dashed lines mark the $3~\sigma$ range used to define the
selection boxes $A, B_{1}$ and $B_{2}$ (see \S \ref{minfrac}).}
\label{box}
\end{figure}

The $MS~sample$ and the $binary~sample$ have been selected according to the
location of stars in the CMD.
We first defined a $V$ magnitude range that extends from 1 to 4 magnitudes below the 
cluster turn-off. In this magnitude range the completeness factor is always
$\phi>$50\% (see Sect. \ref{blend}).
The extremes of the adopted magnitude range ($V_{up}$ and $V_{down}$) have been
converted into masses ($M_{up}$ and $M_{down}$) using the mass-luminosity
relation of Pietrinferni et al. (2006)\footnote{We assumed the metallicities, the distance moduli 
and reddening coefficients listed in the WEBDA database and the extintion
coefficient by Savage \& Mathis (1979). 
Small shifts in the distance moduli ($\Delta (m-M)_{0}<0.1$) have been applied 
in order to match the overall MS-TO shape.}. 
Then, we defined three regions of the CMD (see Fig. \ref{box}) as follows:
\begin{itemize}
\item{A region (A) containing all stars with $V_{down}<V<V_{up}$ and a 
color difference from the MS mean ridge line
smaller then 3 times the color photometric error corresponding to their magnitude
(dark grey area in Fig. \ref{box}). This area contains all the single MS stars in the above magnitude 
range and binary systems with $q<q_{min}$;}
\item{We calculated the location in the CMD of a binary system formed by a
primary star with mass
$M_{up}$ (and $M_{down}$ respectively) and different mass-ratios $q$ ranging from 0 to 1. 
These two tracks connect the MS mean ridge line with the equal mass binary 
sequence (which is $\sim$0.752 mag
brighter than the MS ridge line) defining an area ($B_{1}$) in the CMD. 
This area contains all the binary systems with $q<1$ and whose primary component 
has a mass $M_{down}<M_{1}<M_{up}$;}
\item{A region ($B_{2}$) containing all stars with magnitude
$V_{down}-0.752<V<V_{up}-0.752$ and whose color difference from the 
equal mass binary sequence 
is comprised between zero and 3 times the color photometric error corresponding to 
their magnitude. This area is populated by binary systems with $q\sim1$ that are shifted to the 
red side of the equal-mass binary sequences because of photometric errors;}
\end{itemize}

We considered single $MS~sample$ objects all stars contained in A,
$binary~sample$ objects all stars contained in $B_{1}$ and
$B_{2}$ but not in A (grey area in Fig. \ref{box});

\subsection{Artificial stars}
\label{blend}

Artificial star experiments are essential for a correct estimate of the binary
fraction to account for the effect of blend contamination, photometric errors
and incompleteness on the distribution of single and binary stars in the CMD 
(Bellazzini et al. 2002).

\begin{figure}
 \includegraphics[width=8.7cm]{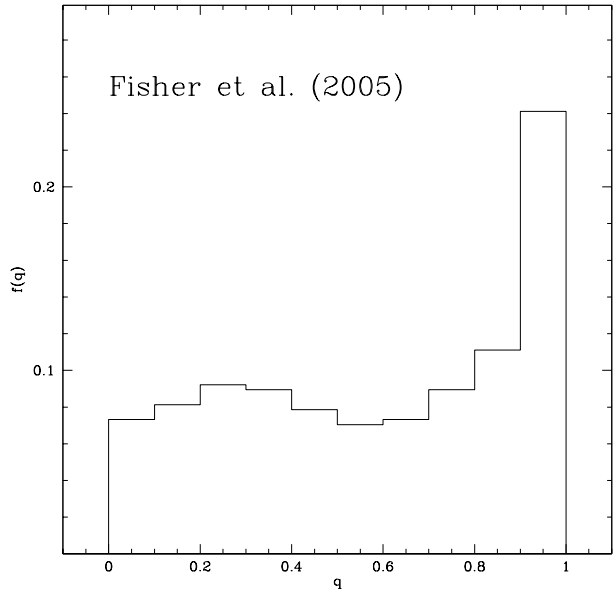}
\caption{Distribution of mass-ratios adopted from Fisher et al. (2005).}
\label{fq}
\end{figure}

For each individual cluster we simulated a population of synthetic single and
binary stars adopting the following procedure.
\begin{itemize}
\item{The masses of $\sim$ 100,000 artificial single stars have been randomly extracted from a De
Marchi et al. (2005) Initial Mass Function (IMF) and converted in V magnitudes (see
Sect. \ref{box_sec}). 
The color of each star has been obtained by deriving, for each 
extracted V magnitude, the corresponding B magnitude by interpolating on the 
cluster ridge line. Thus, all the artificial stars lie on the cluster ridge line;}
\item{For the binary population, mass-ratios have been extracted from the
distribution $f(q)$ by Fisher et al. (2005) shown in Fig. \ref{fq}. 
Then, for each binary system the mass of the 
primary component has been extracted from a De Marchi et al. (2005) IMF and the 
mass of the secondary component has been calculated. The $B$ and $V$ magnitudes 
of the two components have been therefore derived and the corresponding fluxes
have been 
summed in order to obtain the $B$ and $V$ magnitudes of the unresolved binary 
system;}
\item{We divided the frames into grids of cells of known width (30 pixels) and randomly positioned 
only one artificial star per cell for each run\footnote{We constrain each artificial star to 
have a minimum distance (5 pixels) from the edges of the cell. In this way we can control the 
minimum distance between adjacent artificial stars. At each run the absolute position of the 
grid is randomly changed in a way that, after a large number of experiments, the stars are 
uniformly distributed in coordinates. Given the small stars density in the analysed cluster areas, 
the radial dependence of the completeness factor turns of to be neglegible.};}
\item{Artificial stars have been simulated and added on the original frames including Poisson photon noise. 
Each star has been added to both B and V frames. 
The measurement process has been therefore repeated adopting the same procedure of the 
original measures.}
\end{itemize}

This procedure provides a robust estimate of the blending
contamination together with the levels of
photometric accuracy and completeness in all the regions of the CMD and
throughout the cluster extension. 

\subsection{Field stars}
\label{field}

Another potentially important contamination effect is due to the presence of background and foreground field 
stars that contaminate the binary region of the CMD.
This is particularly important in OCs which contains a small number of stars and
are generally located closer to the Galactic plane with respect to GCs.

Fortunately, our observations cover in most cases the entire cluster extension,
providing a good sampling of the field population surrounding each cluster.
As a further check, we used the Galactic model of Robin et al. (2003). A
catalog covering an area of 1 square degree around each cluster center has been retrived.
After scaling for the different field of view, the number of field stars estimated by
the two approaches turns out to be very similar ($\Delta
N_{field}/N_{field}<10$\%). According to both methods the number of field stars in the 
core of all the OCs of our sample never exceeds 10.

Therefore, whenever possible, we used as reference field CMD the one obtained in
the external region of our images. 
For NGC2243, whose observations covers only a small fraction of the cluster
extent, the Galactic model of Robin et al. (2003) has been used.
In this last case, each synthetic field star has been added as an artificial
star to the original $B$ and $V$ frames and the photometric analysis has been
performed. This task accounts for the effects of incompleteness, 
photometric errors and blending. 
 
\section{Estimates of the Binary fraction}
\label{frac}

\subsection{The minimum binary fraction}
\label{minfrac}

In this section we describe the adopted approach to estimate the fraction of 
binaries with $q>q_{min}$. This quantity represents a lower limit to the 
{\it complete} cluster binary fraction.

To derive an accurate estimate of this quantity we simply assumed that {\it all}
the objects of the $MS~sample$ are single MS stars and {\it only} the objects of
the $binary~sample$ are binary stars.
This assumption is equivalent to assume that all binary systems in the cluster
have $q>q_{min}$.

Since the selection boxes defined above cover two different 
regions of the CMD with different
completeness levels, we assigned to each star lying in the $MS~sample$ and in
the $binary~sample$ a completeness factor $c_{i}$ according to its magnitude 
(Bailyn et al. 1992). 
Then, the corrected number of stars in each sample 
($N_{MS}^{obs}$ and $N_{bin}^{obs}$) has been calculated as
$$N=\sum_{i} \frac{1}{c_{i}}$$

We repeated the same procedure for the samples of artificial single
stars and field stars,
obtaining the quantities $N_{MS}^{art}$ and $N_{bin}^{art}$ for the 
$artificial~stars~sample$ and
$N_{MS}^{field}$ and $N_{bin}^{field}$ for the $field~stars~sample$;

Then, we calculated the normalization factor $\eta$ for the 
$artificial~stars~sample$ by comparing the
number of stars in the MS selection box
$$\eta=\frac{N_{MS}^{obs}}{N_{MS}^{art}}$$

The minimum binary fraction, corrected for field stars and blended sources, turns out to be
$$
\xi_{min}=\frac{N_{bin}^{obs}-N_{bin}^{field}-\eta~N_{bin}^{art}}{(N_{MS}^{obs}-N_{MS}^{field})+(N_{bin}^{obs}-N_{bin}^{field}-\eta~N_{bin}^{art})}
$$

The procedure described above has been conducted considering only cluster stars (and
artificial stars) located inside one core radius (see Sect. \ref{prof_sec}). 

The obtained minimum binary fractions $\xi_{min}$ for the clusters in our sample are listed in
Table 2. The typical error (calculated
by taking into account of the Poisson statistic and the uncertainties in the 
completeness corrections) is of the order of 5\%. As can be noted, the minimum binary fraction $\xi_{min}$ is 
larger than 11\% in all the clusters of our sample. 

\begin{table}
 \centering
  \caption{Core binary fractions estimated for the target clusters}
  \begin{tabular}{@{}lcccr@{}}
  \hline
   Name     & $\xi_{min}$ & $\xi_{F}$ & $\sigma_{\xi_{F}}$ & $q_{min}$\\
            &    \%       &  \%       &   \%               &    \\
 \hline
 NGC 188    & 21.0  & 58.2 & 13.5 & 0.52\\
 NGC 2204   & 11.9  & 35.9 & 9.2  & 0.50\\
 NGC 2243   & 34.1  & 70.2 & 9.7  & 0.55\\
 NGC 2420   & 16.6  & 51.4 & 11.3 & 0.52\\
 NGC 2516   & 24.7  & 65.5 & 24.3 & 0.48\\
 \hline
\end{tabular}
\end{table}

\subsection{The complete binary fraction}
\label{true}

The procedure described above allowed us to estimate the minimum 
binary fraction $\xi_{min}$ without any (arbitrary) 
assumption on the distribution of mass-ratios  $f(q)$. However, caution must be used when 
comparing the derived binary fraction among the different clusters of our sample. 
In fact, the definition of the $MS~sample$ and $binary~sample$ given in Sect. \ref{minfrac} depends on the 
photometric accuracy that vary from cluster to cluster.
An alternative approach consists in the assumption of a given distribution 
$f(q)$ and in the comparison between the 
color distribution of simulated stars and the observed CMD.
Until now there are neither theoretical
arguments nor observational constraints to the shape of $f(q)$ in stellar
clusters.
Fisher et al. (2005) estimated the mass-ratio distribution $f(q)$ in the binary 
population of the local field (at distances $d<100~pc$).
They found that most binary
systems are formed by similar mass components ($q\sim 1$). 
Although this distribution is subject to significant observational 
uncertainties and is derived for binary systems in a different environment, 
it represents one of the few observational constraints to $f(q)$ which can 
be found in literature.

In the following we calculate the binary fraction $\xi$ in the target clusters
assuming the distribution $f(q)$ measured by
Fisher et al. (2005, see Fig. \ref{fq}).

To derive this quantity, we simulated a population of $\sim$ 100,000 
artificial single and binary stars (see Sect. \ref{blend}) and calculated the
ratios $\psi_{s}$ and $\psi_{b}$ between the number of $binary~sample$ objects ($N_{bin}$)
and $MS~sample$ objects ($N_{MS}$) for the simulated population of single stars and binaries, respectively.
The quantities $N_{bin}$ and $N_{MS}$ have been therefore
counted in the observed CMD and in the reference field CMD.
The binary fraction that allows to reproduce the observed ratio $\psi_{obs}$ has
been therefore calculated using the formula

$$
\xi_{F}=\frac{(1-\psi_{b})[\psi_{s}(N_{MS}^{obs}-N_{MS}^{field})-(N_{bin}^{obs}-N_{bin}^{field})]}
{(\psi_{s}-\psi_{b})(N_{MS}^{obs}+N_{bin}^{obs}-N_{MS}^{field}-N_{bin}^{field})}
$$

Errors have been calculated according to the standard error propagation and 
assuming a Poisson statistic error on star counts ($\sigma_{N}\simeq\sqrt{N}$).
Of course, the small fraction of stars present in the cluster cores produces
uncertainties as large as 10-15\%.

\begin{figure*}
 \includegraphics[width=12cm]{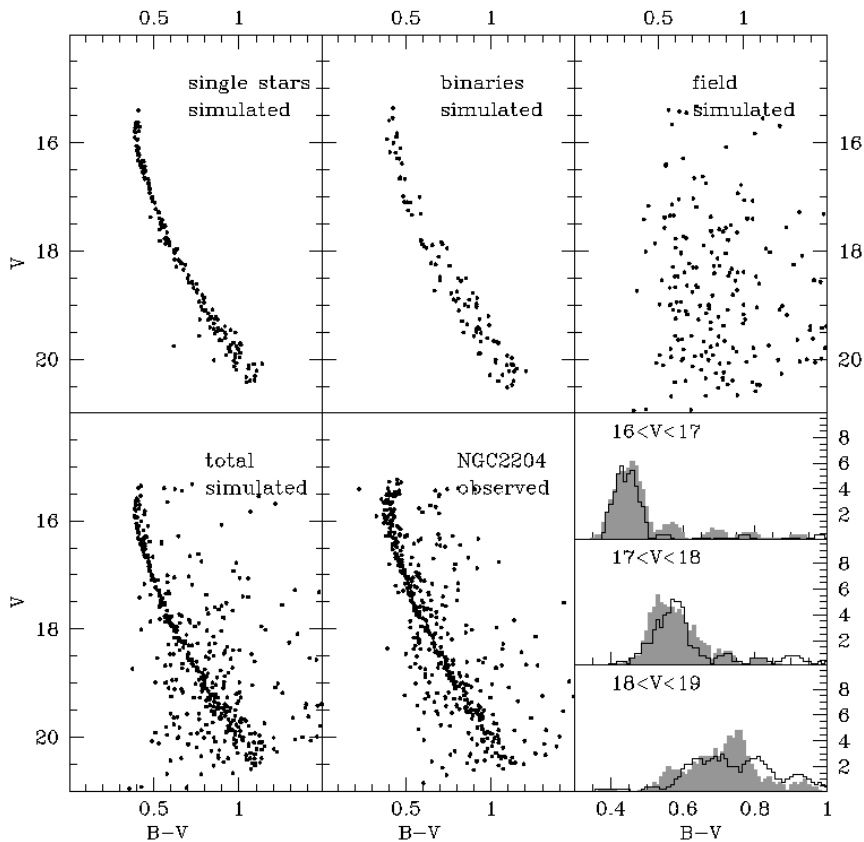}
\caption{Simulated ($lower~left~panel$) and observed ($lower~central~panel$) CMD 
of the core of NGC2204. The color histograms of simulated (black histograms) and observed (grey
histograms) stars in different magnitude ranges are shown in $lower~right~panels$. In the $upper~panels$ the individual CMDs of the simulated single stars
 ($upper~left~panel$), binaries ($upper~central~panel$) and field stars
($upper~right~panel$) are shown.}
\label{sim}
\end{figure*}

A typical outcome of the procedure described above is showed in 
Fig. \ref{sim} where a simulated CMD of NGC2204 is compared with the observed 
one. The overall shape of the observed CMD turns out to be well reproduced. In
particular, the spread of the MS calculated in the observed and in the simulated
CMD between 1 and 4 mag below the cluster turn-off agree within 0.005 mag (see
Fig. \ref{sim}). 
This represents a good quality check to the photometric errors estimates.  

The obtained binary fractions $\xi_{F}$ and their corresponding errors 
are listed in Table 2\footnote{It is worth noting that, because of the different
ages and metallicities of the various clusters, 
the stars belonging to the $binary~sample$ and to the $MS~sample$ span 
slightly different mass ranges in each cluster. However, this effect is taken into
account in the simulated CMDs which have been constructed using a suitable mass-luminosity
relation for each cluster according to its age and metallicity.}. 

As expected, the
values of $\xi_{F}$ estimated following the assumption of a Fisher et al. (2005) 
$f(q)$ are larger than the minimum binary fraction $\xi_{min}$. Note
that neither the ranking nor the relative proportions of the binary fractions estimated
among the different clusters of the sample appear to depend on
the assumption of the shape of $f(q)$.  

For some clusters of our sample the binary fraction were already 
estimated in previous works. Anthony-Twarog et al. (1990), Lee, Kang
\& Ann (1999) Kim et al. (2001) and Sharma et al. (2006) estimated a binary fraction
$40\%
<\xi<50\%
$ for NGC2420 by adopting a technique similar to the one adopted here.
These values agree within the uncertainties with our estimates.
Note that part of the small (although not significant) overestimation of binaries 
predicted in the present work for this cluster with respect to literature values is due 
to the fact that we restricted our analysis to the core of the cluster where the
majority of binaries are expected to be located.  
Minimum binary fractions have been also estimated for NGC2243 ($>$30\%, Bonifazi
et al. 1991) and NGC2516 (26\%, Gonzalez \& Lapasset 2000; 26\%, Jeffries et al.
2001; 40\% Sung et al. 2002). All these estimates are in agreement with the
values derived in the present analysis.

In the following section we compare the obtained binary fractions among the clusters of
our sample and with the sample of GCs presented in Sollima et al. (2007) 
as a function of their main physical parameters. 

\section{Correlations} 
\label{compar}

As introduced in Sect. \ref{intro} the sample of OCs analysed here cover a range in
masses and ages still unexplored by previous analysis on GCs. 
Therefore, the derived binary fractions can be used to check
the validity of the correlations with age and absolute magnitude already noticed
by Sollima et al. (2007) and Milone et al. (2008).

For this purpose we correlated the {\it complete} binary fractions $\xi_{F}$ and $\xi_{min}$ 
with age and visual absolute magnitude and computed the Spearman's rank and Kendall's $\tau$
coefficients for the sample of OCs and GCs separately and for the merged
sample. The results for $\xi_{F}$ are reported in Table 3. 

\begin{figure}
 \includegraphics[width=8.7cm]{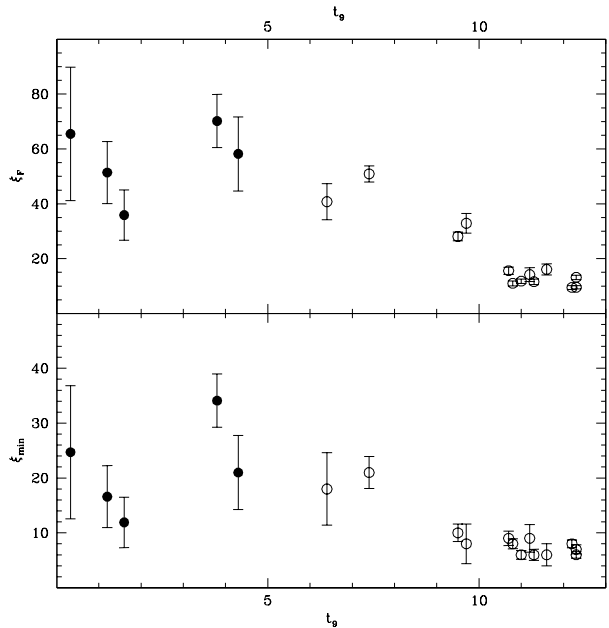}
\caption{Minimum ($lower~panel$), and {\it complete} ($upper~panel$) 
binary fractions as a function of cluster age for the target OCs (filled points) and GCs (open
points).}
\label{age}
\end{figure}

In Fig. \ref{age} the core binary fractions $\xi_{min}$ and $\xi_{F}$ are 
plotted as a function of the clusters age ($t_{9}$). 
As can be seen, a clear anticorrelation between $\xi$ and $t_{9}$ is
evident among GCs (as already reported by Sollima et al. 2007).
Both correlation tests indicates indeed high probabilities within the GCs
sample. 
OCs, which are sistematically younger than GCs, have on average higher 
fraction of binaries.
Nevertheless, within the sample of OCs analysed here, the fraction of binaries
seems to be rather independent on age, as indicated by the performed statistical tests.
Unfortunately, the large errors on the measured binary fractions in OCs make
difficult any firm conclusion on this issue.

\begin{figure}
 \includegraphics[width=8.7cm]{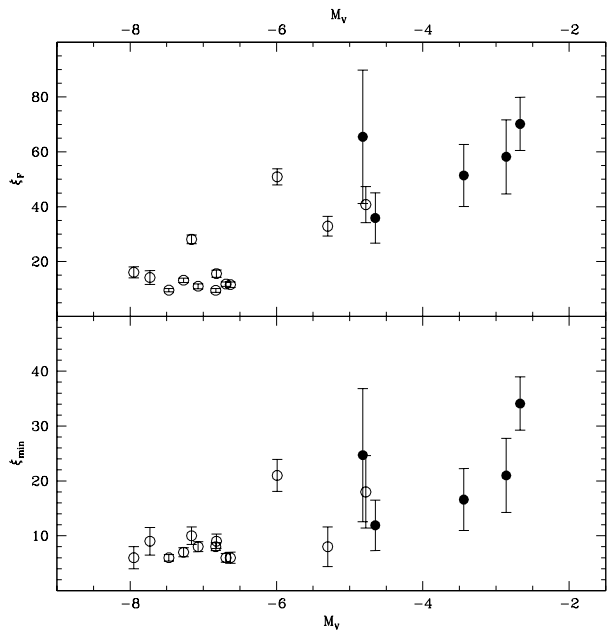}
\caption{Minimum ($lower~panel$), and {\it complete} ($upper~panel$) 
binary fractions as a function of cluster absolute V magnitude for the target OCs 
(filled points) and GCs (open points).}
\label{mv}
\end{figure}

In Fig. \ref{mv} the core binary fractions $\xi_{min}$ and $\xi_{F}$ are 
plotted as a function of the clusters visual absolute magnitude ($M_{V}$). 
This quantity represents the observational counterpart of the cluster mass.
In this case, a nice correlation is visible among both GCs and OCs.
According to the two performed statistical tests, such a correlation seems 
to be significant in the OCs sample and in the same direction for both 
OCs and GCs samples.
The statistical significances of the Spearman's rank and Kendall's tau 
correlation tests on the merged sample are 76\% and 99.8\%, respectively.

We adopted the same procedure to test the correlations with other general and 
structural parameters (metallicity, concentration, central density, core radius,
ecc.). No other significant correlators with $\xi_{F}$ or $\xi_{min}$ have been found among
these parameters. 

Summarizing, the above analysis indicates that the mass seems to
be a good correlator with the binary fraction in both type of stellar
systems. A possible dependence of the binary fraction on age cannot be excluded.

\begin{table}
 \centering
  \caption{Spearman's rank ($\rho$) and Kendall's $\tau$ correlation coefficients for the clusters of
  our sample}
  \begin{tabular}{@{}lcccccr@{}}
  \hline
Parameter & \multicolumn{2}{c}{OCs} & \multicolumn{2}{c}{GCs} & \multicolumn{2}{c}{All}\\
          & $\rho$     &     $\tau$ & $\rho$         & $\tau$ & $\rho$ & $\tau$ \\
 \hline
$t_{9}$   &  0.100 & 0.000 & -0.775 & -0.590 & -0.870 & -0.686\\
$M_{V}$   &  0.400 & 0.400 &  0.390 &  0.231 & 0.734  & 0.529\\
 \hline
\end{tabular}
\end{table}

\section{Discussion}

In this paper we analysed the core binary population of five 
Galactic OCs with the aim of studying their frequency in stellar systems.  

In all the analysed clusters the minimum binary fraction contained 
within one core radius is greater than 11\%. This quantity seems to represent a 
lower limit to the binary fraction in OCs. 
The existing estimates of the binary fraction in other nearby OCs from radial velocity
surveys agree with this lower limit (Mermilliod, Grenon \& Mayor 2008; Mermilliod,
Queloz \& Mayor 2008; Mermilliod et al. 2008). 

The {\it complete} fraction of binaries obtained in this paper range from 35\% to 70\%.
According to the theoretical N-body simulations of Portegies-Zwart et al. (2004) the
binary fraction in a stellar system with a mass of $M=1600~M_{\odot}$ and a
distance to the Galactic center of $d=12.1~Kpc$ (i.e. a typical OC) would
remain actually constant during the entire cluster evolution, rising in the
cluster core from an initial 50\% up to 70-80\% in 1 Gyr of evolution as a 
result of mass segregation.
The fraction measured here are marginally smaller than the final result of 
these simulations.
Following these considerations, the initial binary fraction in our target 
clusters could be $\sim30-50$\%, comparable to that observed in the solar 
neighborhood (Abt \& Levy 1976; Duquennoy \& Mayor 1991; Reid \& Gizis 1997).

The estimated binary fractions of the
clusters of our sample and those of the sample of GCs presented in Sollima et al.
(2007) reveal a significant correlation with the
absolute visual magnitude and cannot exclude a dependence on the cluster age. 
The possible correlation with age found in Sollima et al.
(2007) is indeed not clearly visible among the sample of OCs.
The OCs analysed here hold binary fractions which are on average larger than
those measured in GCs.
However, the small number of OCs and the large errors in the binary fraction
estimates make difficult to assess the existence (or absence) of a
smooth correlation between these two parameters.
A homogeneous analysis of a larger sample of clusters spanning a wide range of
ages is therefore required to clarify this issue.
On the other hand, the correlation between binary fraction and luminosity, already reported by
Milone et al. (2008) in a sample of 50 GCs, has been confirmed also in the
range of mass covered by our sample of OCs (100-1000$M_{\odot}$). 

Theoretical models of evolution of stellar systems with a population of 
primordial binaries (Sollima 2008) indicates that the efficency of the binary
ionization process,
which represents together with mass segregation the dominant process in
determining the binary fraction, has the same dependence as the mass on the 
cluster density and velocity dispersion. 
An anticorrelation between these two quantities is therefore expected.
In practice, in low-mass clusters, a larger fraction of binary systems can
survive to the process of binary destruction as a consequence of the lower rate
of collisions and the smaller mean kinetic energy of colliding
stars\footnote{This could not be the case of NGC2516, whose age is comparable to
(or smaller than) the cluster relaxation time.}.  
The results presented here confirm the prediction of theoretical models
indicating that the process of binary ionization is the dominant
process which determines the fraction of binaries in any relaxed stellar system.
   
\section*{acknowledgements}
This research was supported by the Instituto de Astrofisica de Canarias. 
We thank the anonymous referee for his/her helpful comments and suggestions.

\label{lastpage}

\end{document}